# Genetic information, physical interpreters and thermodynamics; the material-informatic basis of biosemiosis

Peter R. Wills

Integrative Transkriptomik, Zentrum für Bioinformatik Tübingen,

Wilhelm-Schickard-Institut für Informatik, Universität Tübingen, Sand 14, Tübingen 72076, Germany

Department of Physics, The University of Auckland, Private Bag 92019, Auckland 1142, New Zealand

The sequence of nucleotide bases occurring in an organism's DNA is often regarded as a codescript for its construction. However, information in a DNA sequence can only be regarded as a codescript relative to an operational biochemical machine, which the information constrains in such a way as to direct the process of construction. In reality, any biochemical machine for which a DNA codescript is efficacious is itself produced through the mechanical interpretation of an identical or very similar codescript. In these terms the origin of life can be described as a bootstrap process involving the simultaneous accumulation of genetic information and the generation of a machine that interprets it as instructions for its own construction. This problem is discussed within the theoretical frameworks of thermodynamics, informatics and self-reproducing automata, paying special attention to the physico-chemical origin of genetic coding and the conditions, both thermodynamic and informatic, which a system must fulfil in order for it to sustain semiosis. The origin of life is equated with biosemiosis.

Introduction

One of the most vexing questions in biology and the philosophy of biology is the character of information and its relevance to our understanding of the nature of living systems as physico-chemical entities. Much of the vexation is caused by the multi-faceted character of information, not least its relationship to "meaning", of which we have immediate experience in language and the perpetual confrontation of multiple interpretations of even the simplest perceptions. If biosemiotics has anything general to say about this confused state of affairs, it must be that the manifestation of meaning in the physical universe is inextricably linked to biology. However, there is little agreement, except within alternative schools of biosemiotic thinking, as to how that link is to be understood. This paper is not intended to provide the definitive solution to that problem. Rather, a critical analysis of it will be attempted from the point of view of the theories of molecular biology, physics and complex systems, especially as they have been applied to the question of the origin of life. A heuristic approach will be taken, but not one in which any vague notion is tested for its usefulness. Instead, terms with rigorous, formal definitions will be tested for their relevance, with a view to clarifying the standing of physical, computational and semiotic theory in biology. This investigation calls to account the adequacy of current scientific thinking *vis-à-vis* the concrete phenomena of biology, insofar as we are able to describe them by using well-defined terms from different disciplines.

The feature of organisms that distinguishes them most strikingly from everything non-living is the process of hereditary reproduction. Organisms are able to breed true, one generation after another, because their characteristics are, at least in part, encoded in chromosomal DNA that somehow carries a wide variety of information about them. This is true to the extent that a large animal can be cloned simply by transferring a nucleus containing its chromosomes to a denucleated zygote of the chosen species (Wilmut *et al.*, 1997). In this way, an apparently accurate copy of the DNA-donor animal can be grown virtually from scratch. We understand that this is possible because the chromosomal DNA "contain[s] in some kind of code-script the entire pattern of the individual's future development and of its functioning in the mature state" (Schrödinger, 1944). When the base-paired heteropolymeric structure of DNA was first elucidated (Watson and Crick, 1953), Delbrück's idea of hereditary information being stored in a form resembling an "aperiodic crystal" (Timoféeff-Ressovsky *et al.*, 1935) was vindicated. Thereafter certain molecular moieties, first nucleotide bases and then amino acids, came to be conceived of as letters of physico-chemical alphabets and it quickly became apparent that all organisms operate a simple molecular code whereby specific sequences of nucleotide triplets are translated into specific sequences of amino acids comprising the primary linear structure of proteins.

A vivid demonstration of the naturally symbolic function of DNA was recently provided by Gibson *et al.* (2010) who chemically synthesized DNA molecules with sequences closely resembling those normally found in *Mycoplasma mycoides* cells and substituted them for those in a *M. capricolum* cell, after which the transformed cells bred true, through many cycles of reproduction, as a form of *M. mycoides* (the DNA-donor cell) rather than the *M. capricolum* cells from which they had been derived. The identity of the cells changed according to the source of the information in the DNA with which they had been supplied. The DNA synthesis was conducted as an exercise in information processing – the copying and assembly of a suite of nucleotide sequences found in cells of *M. mycoides*, much of the process being controlled electronically. At one stage it was found that an error corresponding to a single nucleotide letter in the synthetic process rendered the transfected cells non-viable and therefore had to be corrected to render molecular biological competence to the cells. Furthermore, the suite of synthesized DNA sequences included some with completely arbitrary base sequences, which are not found in *M. mycoides* or *M. capricolum*, and which would not be expected to serve any predictable biological function. However, when translated purely symbolically using a chosen Roman-alphabet interpretation of base-triplets (Shirriff, 2010), these extra DNA sequences spell out various messages in English: the names of the scientists who worked on the project, an email address, HTML code for reaching a JCVI website, as well as quotations from James Joyce, Richard Feynman and *American Prometheus, a* biography of Robert Oppenheimer intended as an aggrandizing commentary on the significance of the project (Saenz, 2010).

Since the dawn of molecular biology, it has been accepted that the finely differentiated characteristics that distinguish one organism from another are exquisitely dependent on the detailed sequences, sometimes down to the last nucleotide base or amino acid, of the macromolecules found within their cells. Soon after the functionally determinative features (nucleotide base-pairing and sequence) of the genetic material (DNA) and the ribosomal translation of nucleotide sequences into the amino acid sequences of proteins (genetic code) were first discovered, Crick (1958) enunciated two principles to provide a universal molecular explanation of biological specificity: the Sequence Hypothesis and the Central Dogma. While it is true that aspects of both of these principles have required modification, the framework of thought that they established still serves as the *lingua franca* of modern molecular biology and derivative disciplines like bioinfomatics.  Thus, while claiming the heritage of atomic physicists and their empirical analysis of the purely material aspect of natural processes, soon after its inception as a scientific discipline molecular biology departed from a discourse that relied solely on descriptions of the physico-chemical properties of molecules and admitted, in addition, that abstract, immaterial features of components of cells, namely, sequences of letters – chemical moieties construed as hieroglyphics – were causes of biological phenomena.

Although the theoretical perspective of physics and chemistry (blind quantum mechanics) has been radically modified in molecular biology through introduction of the idea of antecedent informational control and processing at the level of single molecular events, there remains a general reluctance among practitioners of the discipline either to admit that such modification is of anything other than merely formal significance, or to consider it as a theoretical problem, except perhaps as a philosophical dilemma of no operational relevance.  Thus, the new discipline of bioinformatics has developed as a branch of information theory and processing devoid of any necessary connection with molecular processes except that the symbols being manipulated entail implicit chemical representations.  What was intended in the original definition of the term *bioinformatics*, "the study of informatic processes in biotic systems ... [an] approach typically involv[ing] spatial, multi-leveled models with many interacting entities whose behaviour is determined by local information" (Hogeweg, 1978; 2011) has been forgotten, save in some theoretical work that is far from the mainstream of molecular biology.

By failing to articulate the general theoretical problem of the relationship between material causation and meaningful information, molecular biology has lost its status as a empirical branch of natural philosophy and has degenerated largely into a set of techniques for determining how biological systems can be manipulated for commercial gain, curiosity or other ends, taking little account of the coincidence of molecular-level physical and informatic processes that differentiate the phenomena of life from everything else found in nature.  Thus, molecular biology has become largely indistinguishable from biotechnology.  While the recent elevation of technology above science by global society is by no means restricted to molecular biology (Forman, 2010) it is of particular significance in that field simply because systems lacking an essentially biological aspect cannot be construed as having any capacity to produce technology.  Neither simple matter nor any simple non-living material object carries the sort of autonomous functionality with respect to which some external structure could be construed as being actively used by it.  However, very basic forms of the capacity for technology are evident in the emergent functional relationships among molecular components of the primitive prebiotic systems from which terrestrial life first arose.  Thus, a theoretically critical molecular biology could, based on the detailed empirical analysis of natural processes, contribute a great deal to the development of fundamental concepts needed for a responsible understanding of the planetary transformations which *H. sapiens* is enabling with the capacities of many new technologies, among them biotechnologies such as genetic engineering (Berg *et al.*, 1975), artificial life (Thiel *et al.*, 2003), synthetic biology (Gutman, 2011) and living technology (Bedau *et al.*, 2010; Wills *et al.*, 2013).

The crux of the problem is the lack of rigorous semiotic theory that can be applied, in an evidently enlightening manner, to the description of molecular events.  When can some atomic-level physical configuration be said to have acquired the function of a sign and how can that occur naturally (as opposed to it being artificially imposed – "designed")?  What conditions must be fulfilled before signs can exist in the physical universe?  I do not presume to know exactly how such a theory can be constructed to answer these questions. However, I will attempt to outline as precisely and accurately as possible the context which must either be accepted as a natural scientific starting point or overthrown in favour of some alternative paradigm. I am not averse to the latter, but it seems worthwhile to understand what parts of current "knowledge" from the natural sciences are deemed to be inadequate, erroneously construed or downright false before everything is discarded and we launch into a programme of untestable speculation that has no grounding in what science since Galileo seems to have placed beyond dispute. It would be desirable for a new theory to make predictions that either bring to light new phenomena or provide for a decisive test relative to extant theory.

The body of the paper is set out as follows. After a discussion of the definition of information commonly used in the quantitative natural sciences, a view of the circumstances under which such information can have meaning is articulated in a sequence of four theses:

1. *The existence of disembodied (non-physical) information about the physical world is incompatible with the second law of thermodynamics.*

2. *Any body of information can be given any meaning whatsoever, by creating a device which functions as an interpreter to deliver the specified meaning upon reception of that information.*

3. *The spontaneous emergence of the algorithmically meaningful interpretation of information is only possible in specialised non-equilibrium physical systems whose components have structures which satisfy certain constraints of a purely formal character.*

4. *Biological evolution consists in the emergent generation and development of autonomous triadic {information, interpreter, meaning}-systems, in which the interpreter is comprised of components from its own output.*

The conclusion reached is that both life and meaning originate in physical systems through a process of information-based self-construction from functionally differentiated component parts.

Information and thermodynamics

Information was first given a rigorous, quantitative definition by Shannon (1948). Although he was concerned with the faithful communication of differentiable patterns of signals, his information measure can be applied to any set of differentiable patterns. The amount of information $H$ produced by a process that has $n$ separately recognisable outcomes, each of which has an independent probability $p_i$, is given by the formula

$$H = -K \sum_{i=1}^{n} p_i \log p_i$$

The constant $K$ specifies the units in which $H$ is measured: if the logarithm is taken to base 2, then $K = 1$ and the unit of measurement as the *shannon* (Battail, 2013), which corresponds to a single toss of a fair two-sided coin. Thus, any defined set of events, measurements, states, symbols or any other possible real-world outcomes whose occurrence is characterized by a discrete probability distribution, is the potential source of an amount of formless information *H*. [The definition is easily extended to continuous probability distributions.] The reasons for adding the qualification "formless" to this measure of information are (i) that it relates to a set of numerical probabilities and has no necessary relation to anything having a material form; and (ii) that any set of abstract forms or qualitative patterns of which an actual distribution of outcomes may be an instantiation, and which could have some contextual meaning, is lost in the reduction to a single numerical value that specifies nothing but the quantity of information. The formlessness of Shannon information can be compared with the formlessness of energy, conceived as a measure of the amount of the most basic substance of which everything in the physical universe is necessarily comprised. Energy has had the status of the formless *Urstoff* of the cosmos ever since Einstein derived his celebrated relation, $E = mc^2$, in which he demonstrated the quantitative equivalence of energy and mass, the classical measure, previously identified with weight, of the amount of substance comprising anything in the physical universe, whether divisible (compound) or indivisible (atomic).

Shannon information begs the question of its definition, "Why *information*?"[1] Answering this question requires that we consider the relevance of Shannon's measure to our more familiar experience of the association of information with representation and meaning. Tossing a fair coin has two possible outcomes whose representation requires just one binary digit, or *bit*, whose values are usually specified as the set {0, 1} (Battail, 2013). The two outcomes have an equal probability of 1/2, and so each toss produces one shannon of information about an actual event in the real world. Any two distinguishable symbols could be used to represent "heads" and "tails" and these symbols would then constitute the bit-alphabet in which the information was represented. Then the meaning of the information, the fact that it represents the outcome of coin tosses, depends completely on some arbitrary historical event or convention whereby the information is associated with those real-world outcomes. Tossing the coin twice has four distinguishable outcomes and produces two shannons of information, and so on. In each case the information in question is a formulaic representation of the qualitative pattern of a particular (set of) outcome(s) whereby it is distinguished from any other (set of) outcome(s): for example, heads rather than tails for one toss; two heads rather than any of the other three possible outcomes for two tosses, and so on. This differentiation of the pattern of individual outcomes, with reference to which the probability distribution over members of a set of outcomes is specified, is the manner in which information has form in any particular context. The quantity of information given by Shannon's formula specifies the overall degree of differentiation between outcomes, but it is silent concerning (i) the manner of differentiation; (ii) the number of possible outcomes associated with the relevant probability distribution; and (iii) details of the range of possibilities in the set of outcomes under consideration, or more significantly, their consequences.

It is possible for information to be produced in the real world only because the physical universe is not homogeneous and not in thermal equilibrium. If the universe were entirely homogeneous and in thermal equilibrium, its temperature and composition would be identical at every location and no information could be produced, because every process would have the same outcome – no change from the initial condition. A process that has only one possible outcome produces no information. However, in this instance we have used the term "process" rather loosely; no thermodynamic processes occur in a system at equilibrium, so there is no probability distribution of outcomes to which Shannon's formula could apply, except the instantaneous occurrence of maximally disordered molecular movements and collisions. The production of information envisaged by Shannon requires the sampling of the probability distribution in question, the determination of something temporally real from among a set of possibilities – an actual occurrence. [2] The probability distribution can be sampled by any system for which each defined outcome of the process under consideration has some distinguishable, consequential effect. In other words, the production of information through the sampling of a set of possible outcomes requires a physical interaction between the system producing the outcomes and the system that is said to do the sampling.

---

[1] There are alternatives to this probabilistic definition of Shannon information (Muller, 2007), most importantly the algorithmic definition which measures the minimum number of bits needed to specify a procedure for generating a given string of letters.

Szilard (1929) and Brouillon (1956) gave detailed consideration to the thermodynamics of the necessary interaction between the information-producing process ("system") and the sampling system ("observer") and were thereby able to exorcise Maxwell's Demon from the physical universe. Maxwell's Demon was a hypothetical entity which could, through the judicious use of information gained by sampling the local fluctuations (microscopic inhomogeneities) that occur in a system at thermal equilibrium, take heat from a cold region of the system and deliver it to a warm region without doing work and then use the temperature difference to do useful work at no net thermal cost. The significance of Szilard and Brouillon's results, and all who have followed them, is that both the production and utilization of information, conceived in any scientific way, are bound to events in the physical world and the laws that govern it. Any purely sentient entity able to gain, through non-interactive observation, the information needed to construct a sufficiently accurate representation of the microscopic details of a physical system would be able to use the mechanism of Maxwell's Demon to create a perpetual motion machine in direct violation of the second law of thermodynamics. In other words, as we currently know them, the laws of physics, especially the second law of thermodynamics, would no longer hold universally if information-generating observation of a material system could be carried out without any physical connection between the observer and the system under observation. The existence of Descartes' immaterial *res cogitans* is incompatible with the thermodynamics of his *res extensa*.

Meaning

Having determined that information whose production was devoid of causal connection to the real-world outcomes it represented would violate physical law, we now have the task of determining what sort of causal connection between real world outcomes and their representations are able to make information meaningful in any way. Perhaps surprisingly, we quickly reach the conclusion that virtually any sort of causal connection between information and its representation is possible – in which case any body of information can have practically any chosen meaning.

Take the information contained in the human genome as an example. It comprises some $6 \times 10^9$ bits. When the physical form of this information consists of a concatenated sequence of nucleotide bases chosen from the canonical four-letter alphabet, that is, comprising DNA molecules as found in human chromosomes, an obvious real-world interpretation of this information is the specification of a particular specimen of the biological genus *H. sapiens*. In the case of the standard sequence published by the Celera Corporation (Venter *et al.*, 2001) it turns out that the human being so specified would be a near clone of the individual known internationally as "J Craig Venter" rather than an individual corresponding to a consensus sequence drawn from a wide range of unique specimens of *H. sapiens* (Levy *et al.*, 2007). There is no reason to believe that humans cannot be cloned in the way other animals have been (Wilmut, 1997), in which case the means already exist for the production of an individual human being as a particular interpretation of some selected body of information of size $\sim 6 \times 10^9$ bits embedded in DNA sequences. However, there exists a multitude of other interpretations of the information in the human genome, the range of which is restricted only by the ingenuity of whoever wishes to build a device capable of giving effect to their potentially perverse imaginings.

---

[2] I refrain from identifying "sampling" with observation in order to avoid associating it with any notion of sentience.

Practitioners of bioinformatics are much more accustomed to dealing with genomic information that is stored in digital electronic form rather than in DNA sequences. That is the whole point of genome sequencing: to represent the information in some easily accessible and manipulable form, averting the need to deal continually with particular DNA sequences in particular locations. This transformation of the storage medium brings new possibilities to light. When information is stored in particular structures it becomes readily available for use in the service of any purpose whose fulfilment depends on processes that makes reference to that form of information. It was this aspect of information that the Venter Institute took advantage of when they encrypted messages into the DNA sequence of their engineered strain of *M. mycoides*. Anyone who finds a cell of that strain, sequences its genome, and chances upon a specific mapping of nucleotide triplets onto an ASCII-like character set will discover, within the otherwise unintelligible sequences of letters, messages written in English. The connection between the meaning of the English sentences and the vain purpose expressed by the Venter Institute in their encrypting those particular sentences into the genome of an organism is a question of "interpretation" far beyond the scope of this paper. What is of interest to us here is the demonstration that there exists an arbitrary information-processing algorithm (the mapping from trinucleotides to ASCII-like character set) whereby part of the genome of an organism can be interpreted as English language messages relevant to certain societal activities. Whether the particular DNA sequence came into existence by accident or design, it would still be open to interpretation as messages in English: the transformation of the information into that form requires only the existence of some physical entity, or constellation of entities, that can execute the relevant algorithm. Anyone can use freely available internet resources to execute a similar algorithm (the standard trinucleotide to one-letter representation of amino acids) and look for the occurrence of his or her own name, or any other set of words, in the human genome.

This discussion is not intended to pre-empt argumentation about whether the processes of nature, from the elementary quantum events in the early hot universe to the neurological dynamics of a thinking person's brain, operate in an algorithmic fashion. Algorithms, like information, insofar as they correspond to anything that is bound to physical contingencies, are patterns that are abstracted from descriptions of the real world. They stand as representations of what appear to be mechanisms, often general mechanisms, according to which natural processes are ordered. As Barbieri (2013) points out, mechanisms serve as scientific models in the realm of natural philosophy. Even the most ardent modern reductionists, who seek complete and final knowledge of nature's unity in the form of a uniquely defined quantum field, eschew the idea that the operation of the relevant algorithm is deterministic. All meanings, including beliefs about nature, are irrelevant to the algorithmic operation of such a Laplacian machine, which also defeats any conscious impression that thinking has physical consequences. Exactly to the contrary, others seek creative freedom in the belief that nature's operation is non-algorithmic and is bound ultimately to confound the efforts of all model-building that relies on descriptions of operational mechanisms (Penrose, 1989; Kauffman, 2012).

Possible interpretations

I now wish to state a **strong claim**:

> *Any body of information of non-zero magnitude can be interpreted, and therefore function as a sign for, any meaning whatsoever. The possible meanings of any body of information depend only on the availability of physical interpreters that generate those meanings.*

Given an elaborate enough description of what is naturally possible and what is observed in the physical world, the occurrence (as opposed to the non-occurrence) of a single event, that is a body of information comprised of a single bit, could be interpreted as a sign that God exists, an answer to one of the most vexing philosophical questions that has ever been posed. This example is provided to illustrate that the meaning of information can often be ascribed almost entirely to the internal structure and rules of the interpreter. However, in attempting to understand how systems of meaning are associated with particular physical structures and processes, focussing especially on elementary bio-molecular systems, I wish to restrict application of the word "interpreter" to the description of a physical system for the transformation of information, a way of mapping one pattern of information onto another, what Barbieri (2003; 2013) calls a "code". I prefer to use "code" to refer to the simplest transformations of information or parts of algorithms, just as we refer to the universal genetic code as a mapping from 64 codons onto 20 canonical amino acids and the STOP signal. The molecular biological interpretation of a genetic sequence as the primary structure of a protein, as produced by a ribosomal-centred computer, requires operation of an algorithm involving recognition of an ATG-trinucleotide START signal, followed by repeated translocation and code-table assignment operations, until a STOP signal is recognised and the termination operation is executed. And it is most interesting that the integrity of the code part of the algorithm is not maintained directly by the ribosomal mechanism, but by an independent suite of enzymes known as amino acyl-tRNA synthetases (AARSs), a theme to which we will return.

Looking at arbitrary non-biological meanings of genetic information will give us a deeper appreciation of the biosemiotic processes that have led to the natural construction of the equally arbitrary, but very complex, system of interpretation of genetic information upon which all life depends and which molecular biology continues to bring to light. For that reason, I would like to consider the interpretation of the information in the human genome as the sound produced by a performance of Beethoven's Ninth Symphony. It is easy to construct a machine which takes the information in the human genome and produces an audio file record of such a performance. All one need do is create a bitwise alignment of the two sets of information, the human genome (A) and the audio record (B), perform the "exclusive or" binary operation across the entire length of the alignment to produce their XOR cross (C), and then embed C in a Turing machine that accepts an input stream and XOR crosses it with its internal record of C to produce an output. If A is given as input, then B will be the output. This is because the XOR has the property that if XOR(A, B) = C, then XOR(A, C) = B and XOR(B, C) = A. A Turing machine constructed in this way would appear to have very little use, relying as it does on a system of states dictated by the information contained in C, but its interpretation of the information in the human genome is likely to be quite robust in respect of input errors. One could corrupt the human genome input information to an extent that it became completely meaningless from a biological point of view, incapable, when instantiated as a DNA sequence, of allowing a human cell to maintain life. However the audio output of the symphonic performance generated by the machine would, to the human ear, probably be indistinguishable from the original soundtrack used to produce C.

*First objection*

The most obvious criticism of this Turing machine construction is of the type: "But the machine already contains Beethoven's Ninth in the embedded information (C). The claims about it being an interpretation of the human genome are fraudulent!" In response it might be pointed out that if the symphonic performance is given as input then the genomic sequence needed to specify a human cell is given as output. So, what does the machine contain, Beethoven's Ninth or the human genome? The real question that this objection raises is whether any body of information has an intrinsic, inseparable meaning. We have answered that question with a resounding "no", without precluding the possibility that information can have an origin that is practically inseparable from it. What is important concerning the interpretation of the human genome as the sound of a performance of Beethoven's Ninth, B = XOR(A, C), is that it would not be possible to create such the interpreter that produces the sound of the performance if the Symphony did not already exist. The chance appearance of our hypothetical Turing machine in the universe without the historical occurrence of Beethoven having written the symphony is inconceivable.[3] However, if we consider the complementary interpretation, A = XOR(B, C), obtaining the human genome from a recording of the symphony, then the way in which information can be bound to its origin becomes much clearer. The human genome first has to exist and be known by us for it to be possible for us to construct the C-dependent Turing machine that retrieves it (A) from the symphony (B).

*Second objection*

The second objection to the effect that our Turing machine cannot reasonably be conceived of as an "interpreter" of any kind concerns the complexity of its internal algorithm. We have allowed the interpreting algorithm to be of about the same informational complexity as the intended input (the human genome) and desired output (the sound of a symphonic performance). By following the procedure described, it would be possible to build a special machine to produce the symphony using any chosen suitably-sized body of information as input. However, far from invalidating the machine's status as an interpreter, this objection leads us to ask how complex the natural process (viewed as an algorithm) is that produces a human being when it is given a human genome as input information, as would occur if a human were cloned. The idea that the sequence of the human genome is all that is needed for the construction of a human being, that it has such an intrinsic meaning, is now shown to be untenable. However, it is still believed by many molecular biologists that the human DNA sequence somehow specified the existence of *H. sapiens* as an animal in some Platonic mathematical space of biological forms long before our species ever existed, an idea that has been spelt out in detail by the most elevated exponent of such scientific error, until recently the Simonyi Professor for the Public Understanding of Science at the University of Oxford (Dawkins, 1986, p73). The question of real biological interest is the extent to which the interpreters of genetic information found in diverse organisms possess functionalities that are held in common from any point of view other than the human perception of them. Gibson *et al*. (2010) seem to have demonstrated what would be expected: that if one organism is similar enough to another, its interpreter may be compatible with the other's genome, within the bounds of current techniques for genomic transplantation.

---

[3] The dependence of the origin of Beethoven's Ninth on the pre-existence of the human genome (the fact of Beethoven's humanity) obfuscates the main question and arises only because the symphony is a human artefact and not some other complex object.

*Third objection*

The strongest argument against the validity of designating our Turing machine as an interpreter of any kind, is its lack of generality of function. The XOR(C, J) output, produced by the machine with an input J not resembling either the A and B used to produce C = XOR(A, B), would not be recognisable as either a genomic sequence or an audio record. Although it could be instantiated as either, the output information could hardly be expected to function either biologically or musically, or any other functional domain, except by an accident of inconceivably low probability, a "miracle". Something which acts as an interpreter, even within the restricted definition of the term accepted within computer science, is usually thought of as taking information from one domain, where the information has the status of a sign, or representation, and mapping it directly into another domain, where it comes to be the thing whose existence was encrypted in the form of the sign. An interpreter is a device that constructs objects from their encrypted descriptions. An audio CD player is a general interpreter for the transfer of information from spatial patterns of information on an optically readable disc to corresponding audio "objects", commonly comprised of temporal patterns of voltages and currents in the solenoids of electromagnetically driven speakers.

On this basis, the genome-symphony Turing machine we have described could indeed be used as a general interpreter of a sign J for any information I whatsoever that had been encrypted through the process J = XOR(C, I). One could even conceive of a device incorporating some hidden C being sold for the retrieval of information (I) from encrypted proprietary records (J). The information (I) could be bioinformatic records, music, books, photographs, anything at all that can be represented digitally. The functional operation of the machine would define the convention for the retrieval of the information, just as there are conventions for the further interpretation of digital records as chromosomes or audio signals or whatever.[4]

---

[4] The XOR encryption system described here would be pitifully ineffective against attempts to circumvent it ("break the code"). However, all of the same arguments apply to much more sophisticated systems of encryption. The XOR system has been chosen only for illustrative purposes.

Precursors to semiosis

Having established that a physical system comprising an *interpreter* is required for information ever to have meaning, we now proceed to enquire into the natural origin of interpreters.  The first observation to be made is that most bodies of information of any size are associated in some way with what can only be regarded as cosmically unique events or structures.  In this sense a body of information can be construed as having a "natural" meaning; merely by existing it "represents" or "signifies" the occurrence of its creation.  By the same token, it is possible to create bodies of information, like the cross between the human genome and a soundtrack of Beethoven's Ninth Symphony, in which "references" to more than one essentially unique occurrence or structure are embedded.  Thus, systems of meaning can become very complex, especially when an object that is the product of a functioning interpreter comprises or contributes to a structure which is taken as a source of information by a second interpreter – all the more so when the second interpreter recognises completely different types of structures as bits of information.  However, in biology there are quite simple natural systems of meaning, what Barbieri (2003; 2013) calls "organic codes", whereby information is transferred between well-defined domains in which it is stored ("represented") or functions.  The execution of each organic code relies on the existence of an interpreter which typically has quite special but generic physical structures.  We will soon see this in relation to the genetic code.  The question in theoretical biology that we now want to address is how interpreters, which give specific and detailed meaning to bodies of information as large as the sequence of the human genome, can come into existence from an initial state of complete molecular disorder.  The fact of genetic information's existence finds a reasonably satisfactory explanation in terms of a molecular theory of Darwinian selection (Eigen, 1971; 2013), but the disposition of its *functional meaning*, above and beyond the fact that the information has accumulated, begs an enquiry into the character and emergence of natural systems of interpretation, i.e., what Barbieri (2013) calls "code biology".

At its most elementary level, an interpreter is a dynamic physical system in which information is transferred from one form, or domain, to another.  At its most complex level, a system of interpretation, a "language" perhaps, is not restricted to transfers of information between existing domains and the meanings that can be can expressed through their operation.  Using a language, people can generate new transfers that simultaneously take novel assemblages of input information from diverse domains and produce output in some previously non-existent domain.  That is, a previously non-existent domain of information can be created within the extant physical reality without altering it to any currently perceivable extent.  In the extreme case of the socially constructed meaning of our everyday lives, the interpretation of the information constituting the patterns of differentiable perceptions open to us seems to be completely emancipated from any physical or mechanical determination, except for the apparently incidental fact that it would not arise without the prior existence of our brains.  So, our task is to locate the possibility of meaningful information transfer within the constraints imposed by the structure of physical reality, but not in such a way that physical laws prescribe, dictate or determine what the meaning or interpretation of any body of information should be.  We will do this by asking a question: How can a physical system in which information is transferred from one domain to another arise naturally? What were the first semiotic systems?

We have already seen that information cannot be transferred within a system at thermodynamic equilibrium or, by extension, in a stable steady state.  Therefore, at the very least, some sort of far-from-equilibrium system is required. What first strikes us, is that far-from-equilibrium systems display a wide variety of ordered macroscopic behaviours, the identifying characteristics of which depend on details of the couplings between molecular processes in the system (Glandsdorff and Prigogine, 1971). For example, the system described by Turing (1952) displays macroscopic spatio-temporal patterning as a direct result of coupling between molecular reaction and diffusion processes. In some far-from-equilibrium systems the internal dynamics are coupled to an internal "relaxed" or "inert" molecular state, detailed aspects of the system that affect, but are not changed by, the molecular processes that take place in the system (Pattee, 1972; 1995). Alternative relaxed molecular states of this sort that affect the behaviour of the system in different ways can serve directly as information that the system interprets – the coherent dynamics of the system producing different outcomes from different "initial conditions" supplied in the form of potentially different inert state "inputs". This is, of course, the manner in which heritable DNA sequences serve as information for the construction, maintenance and evolution of organisms.

If we want to understand semiosis in systems that are bound by the inviolable constraints imposed by the physical structure of reality, then a first step might be to investigate the origin of primitive systems of the character we have just described. How can a system whose dynamic behaviour is coupled to integrated polymeric sequence information emerge from processes of randomly occurring uncoupled molecular interactions? The answer to this question, at least as far as we can describe simple model systems in which this sort of behaviour emerges, has some most interesting features that go to the core of how we might define "life" and understand its local origin. The first important result was obtained by Eigen (1971), who examined a system of polymer sequence copying, in which the domains of information and its interpretation were identical. When the fidelity of sequence replication exceeds a defined threshold value, such a system can generate information, in the form of a predominant (or consensus) polymer sequence. This occurs in a homogeneous mixture of molecules simply as a result of Darwinian selection among polymer species that have different rates of reproduction and loss. In describing how polymer sequence information can accumulate, essentially *ex nihilo*, Eigen (1971) assumed that any molecules responsible for the process of copying, e.g., replicase enzymes, were supplied and maintained externally, in no way controlled by the system. In the extended system of the hypercycle (Eigen, 1971; Eigen and Schuster, 1978) the replicases came to be produced within the system and strict Darwinian competition between polymer information carriers was circumvented. The hypercycle demonstrated the possibility of information carriers coexisting when their dependence on one another was arranged in a closed cycle of mutual amplification. However, the hypercycle introduced a new level of interpretation of molecular sequence information, above and beyond the "zeroth order interpretation" entailed in the process of copying. In the hypercycle, the information in each replicating polymer is interpreted, through some unspecified means, to produce a specific replicase enzyme which preferentially catalyses replication of the next polymer, making a directed hyper-connection from one cyclical process (replication of a specified polymeric information carrier) to the next. Hypercyclic systems demonstrate marginal prebiotic functionality, but they beg our original question, which we can now rephrase: How can systems capable of interpreting sequence information emerge as a result of birth-death turnover processes in populations of polymeric macromolecules?

The work of Kauffman (1986; 1993; 1995) is representative of a quite different approach to the problem, one in which any information-carrying capacity of macromolecules is quite secondary to the virtually inevitable emergence of a more basic protobiological order in dynamic populations of polymers. Kauffman's autocatalytic sets of polymers rely on the propensity of molecules like proteins to act as catalysts of specific ligation and/or cleavage reactions, using highly specific recognition processes to select substrates and concatenate them together to form new sequences, which in turn may catalyse other reactions. In its original conception (Kauffman, 1971; 1986) the propensity of proteins for catalysis and their capability for specific substrate recognition were considered to be so high that nearly every possible polymeric sequence could be expected to participate in an enormous network of coincidentally reinforcing cross-catalytic interactions. An alternative approach to the problem (Wills and Henderson, 2000) adopted the implicit view that catalytic polymers should be construed as monadic interpreters that take the substrate sequence motifs they specifically recognize as input, and produce their concatenation as output. An attempt was made to determine realistic limits on the degree of functional specificity that could be ascribed to an interpreter comprised of polymeric structures of given complexity. Model autocatalytic systems based on these principles can only exist by meeting a high degree of coincidence concerning the extended sequence elements from which catalytic interpreters can be built up and the restricted set of elements that exist as "interpretations" generated by the rules of the system.

Eigen (1971; 2013) insists that the level of catalytic specificity typical of biological systems cannot be achieved by self-organisation of autocatalysis alone, rather that the replication of information carriers is necessary. Recent reconsiderations of Kauffman systems (Hordijk, Kauffman & Steel, 2011) have been more realistic than the original (Kauffman, 1986), but take little account of the potentially symbolic character of polymer sequence information. The process of local sequence-complement matching has been incorporated as one constraint on catalysis, but like DNA copying, which is actually a monadic two-step process, it doesn't amount to more than "zeroth order interpretation". Hordijk, Wills & Steel (2013) have attempted to bridge the conflicting positions (Eigen, 1971; 2013; Kauffman, 1971; 1986; Hordijk, Kauffman & Steel, 2011) but only with partial success.

The significance of this gamut of work is that it demonstrates the possibility of the emergent interpretation of polymer sequence information, at least of the zeroth order type, in autocatalytic systems. However, the inherent system-wide control of process and structure specificity is much less than can be achieved is biological systems that rely on polymers that serve as information carriers (Eigen, 1971). Nevertheless, the systems studied can be construed as displaying a primitive form of self-referential "semantic closure" whereby, according to Pattee (1995), "the freely selected symbolic aspects of matter ... [allow] ... the law-determined physical aspects of matter [to] become functional". The structural complexity and functional specificity that can be achieved in such systems, given the dynamic constraints on them, appears to be very low (Wills and Henderson, 2000; Hordijk, Wills & Steel, 2013) and alone they do not serve as an adequate foundation upon which more general systems for the interpretation of molecular sequence information can be built up.

Coding systems

The first direct investigations of the evolution of systems that support the autonomous interpretation of molecular sequence information are due to Bedian (1983) and Wills (1993). The basic assumption made by these authors was that simple chemistry includes the possibility of a second class of polymer (such as a protein) being synthesised in such a way that molecules belonging to some other first class of polymer (like a nucleic acid) act as information-carrying templates that function as "instructions".[5] With the polymer molecule from the first class acting as a template, the second polymer molecule is constructed with a sequence that bears a relationship of collinearity with the sequence of the first polymer. However, a mechanism of collinear synthesis does not of itself require or guarantee the transfer of information from the template to the polymer molecules of the second class. Information transfer requires that the selection of the monomer added to the growing polymer chain at any position be influenced by the identity of the monomer located at the collinear position on the template. By means of such influence, coded information transfer is able to emerge as a self-organised behaviour of the system, generating itself *de novo* from an initial state in which the sequences of the polymers being synthesised are completely random. In this initial state there is no net template-influenced selectivity of the monomer concatenated to the growing polymer chain at any position, because all assignments are equally probable.

Coding self-organisation works as follows. Suppose we are dealing with binary alphabets of amino acids {*a*, *b*} and template sequence elements {*A*, *B*} that act as "codons"; and that two particular proteins (assignment catalyst enzymes) with sequences $E_1$ = *baababbaaab* and $E_2$ = *abbbaababba* are able to catalyse the two assignment functions $\{A \to a\}$ and $\{B \to b\}$ when proteins (amino acid polymer chains) are synthesised in a manner collinear with extant nucleic acid templates (sequences of codons). In a thermodynamically driven system in which proteins are synthesised from energetically activated amino acids *a\** and *b\**, the presence of nucleic acid templates with (arbitrarily chosen[6]) sequences $T_1$ =BAABABBAAAB and $T_2$ = ABBBAABABBA will enable the two coding enzymes to produce themselves autocatalytically. Coding self-organisation occurs when a population of proteins with random sequences and very weak catalytic activity for all of the possible assignments $X \to y$ use the specified templates for synthesising further protein molecules. The newly synthesised population is also random until stochastic fluctuations occur that simultaneously increase the concentrations of $E_1$ and $E_2$ above some critical threshold, after which self-amplification of these two species proceeds until they dominate the protein population and the system operates the coded transfer of individual bits of information according to the rules $\{A \to a\}$ and $\{B \to b\}$.

From the point of view of physics and chemistry there is nothing mysterious about the process of self-organisation in Bedian-Wills systems – it occurs as a result of the inexorable operation of physico-chemical laws. On the other hand it can only occur when there is a particular *formal relationship*, which has nothing to do with energetic or thermodynamic constraints, between the *sequence information* carried by the template species available to the system (e.g., $T_1$ and $T_2$) and the structure-dependent (catalytic) *properties* of molecules of the second polymer species (e.g., $E_1$ and $E_2$). The required relationship is that the information (in $T_1$ and $T_2$) must be "reflexive" in the sense that when it is interpreted according to the rules of an emergent code the second-class polymers that are synthesised must have sequences corresponding to catalysts ($E_1$ and $E_2$) that happen to carry out only the subset of assignments belonging to the code $\{A \to a, B \to b\}$ and not the subset of other (non-code) assignments $\{A \to b, B \to a\}$.[7] The required coincidence of cause and effect is extraordinary and in simulations of coding self-organisation it is set up in advance by the programmer (Bedian 1982; Wills, 1993).[8]

---

[5] The template could be designated as a catalyst because concatenation of monomer units of the second polymer is promoted through interactions with the template, which remains unchanged in the process. However, it is polymers of the second type that are usually considered as catalysts in the system, because they have the capability of catalysing the concatenation of a new monomer unit onto a growing chain of the same the class to which they belong, the way ribosomal proteins are capable of catalysing the transfer of amino acids onto growing oligo-peptide chains.

[6] Reverse-order complementary sequences for $T_1$ and $T_2$ have been chosen for later illustrative purposes.

When the required correlation exists, the possibility of semantic closure (Pattee, 1982; 1995) emerging through the process of coding self-organisation is built into the system as an effect of autocatalysis: polymers of the sort being synthesised play a role in the synthetic process that produces them, in effect serving as adaptors, influencing the choice of monomer to be concatenated to the partially synthesised polymer chain according to the particular monomer present at the collinear position on the template. Coding self-organisation is a selection process in which the population of polymers being synthesised becomes enriched in molecules that catalyse particular monomer assignments, while molecules that catalyse alternative monomer assignments, destructive of the emerging code, progressively disappear from the population. Through appropriate template choice, the required trans-species information-function correlation is a possibility even if the relationship between the structures and catalytic properties of the second polymer species is essentially random. However, the formal correlation between the sequence information of molecules of the template polymers and the structure-determined properties of the catalytic polymers remains as an indispensable *extra-physical* requirement for thermodynamically driven coding self-organisation.

The formal trans-species information-structure correlation that enables the emergence of semantic closure in systems that undergo coding self-organisation can rather easily be attributed to the template sequence alone: the template sequence is required to have the special, and extremely rare, property of "reflexivity" *vis-à-vis* the sequences of catalysts that execute the coding assignments (Wills, 2001). However, adopting this perspective on the origin of coding condemns us to a "miraculous" view of the appearance of template polymers with the necessary sequence properties. Under any physically realistic assumptions, the chance appearance of reflexive template sequences turns out to be far too improbable for it to play any useful role in the scientific explanation of biosemiosis. But the theoretical situation then turns out to be even more dire. If a reflexive template sequence were provided to such a system, it would quickly be lost and replaced by inferior species as a result of errors in its reproduction and the extant coding system would be lost (Wills, 1994). Finally, to top it off, it can be demonstrated analytically that natural selection cannot provide for survival of the required species in an ordinary, spatially homogeneous physical system, even when template reproduction is catalysed by some second-species polymer that is integral to the system (Füchslin and McCaskill, 2001). At this stage it seems hopeless: the theory of coding self-organisation suggests that some sort of continuous physico-chemical miracle, the unlawful material intervention of an immaterial agency, is required for a molecular-level interpreter to be able to exist by maintaining (copying) and interpreting a stored description of its components.

---

[7] The information-function relationship of coding reflexivity is not always unproblematic. It can only be unambiguously established for classes of polymers whose structure-function relationship (mapping from polymer sequence to catalytic properties) is sufficiently asymmetric (Nieselt-Struwe & Wills, 1997).

[8] This is accomp0lished by symbolically "back-translating" the sequences of $E_1$ and $E_2$ to determine the sequences of $T_1$ and $T_2$. A similar procedure must be followed to produce self-reproducing automata (von Neumann, 1966). The "description" tape must be devised and specified by the programmer.

Spatio-temporal self-organisation

The physico-chemical miracle required to save chemical coding systems is supplied by the mechanism described by Turing (1952). Füchslin and McCaskill (2001) were the first to demonstrate that the pessimistic outlook for the theory of coding self-organisation just outlined is the result of taking a too simple, molecular, chemical perspective. If one considers the movement of molecules as well as their chemical structure and properties, and the possibility of coupling between molecular diffusion and chemical reactions, then individual molecular template sequences can become sufficiently localised in the immediate vicinity of limited subsets of second-species polymer catalysts for natural selection to favour the joint survival of reflexive information and the corresponding autocatalytic set of catalysts comprising an interpreter. As noted above, the correspondence between the template information and sequences of the population of catalytic polymers acts as an extra-physical constraint on the joint selection of polymers from the two classes. The systems of this sort considered by Füchslin and McCaskill (2001) and later by Markowitz *et al.* (2006) have been called gene-replicase-translatase (GRT) systems. Their architecture is now under study in the general domain of machine information-processing and learning (Tangen, 2013) and there are efforts afoot to use the main features of GRT systems in the design of electronic-chemical devices (McCaskill *et al.*, 2012).

The behavioural feature of GRT systems that separates them from simpler molecular coding models (Bedian 1982; Wills 1993) is the spatio-temporal self-organisation that occurs as a result of reaction-diffusion coupling, an effect that was first brought to light by Turing (1952) in his consideration of the principles of morphogenesis. Turing's theoretical demonstration of the possibility of spontaneous spatio-temporal self-organisation in reaction-diffusion systems predated the mature development of irreversible thermodynamics, especially the rigorous study of nonlinear systems, and any understanding of the manner in which nonlinearities in system dynamics can give rise to all sorts of "dissipative structures", considered to be so extraordinary when seen from the naïve molecular perspective that they were dubbed "a new state of matter" (Prigogine, 1978). In his classical treatment of reaction-diffusion systems, Turing (1952) described the spontaneous emergence of macroscopic chemical patterns, spatial inhomogeneities, in systems that are initially homogeneous. The result is counter-intuitive, as is the existence of biological systems, given a superficial understanding of the second law of thermodynamics. The slowness of the scientific community to accept the reality of the Belousov-Zhabotinsky reaction attests to this difficulty (Winfree, 1884; Frank-Kamenetskii, 2012).

The theoretical attraction of GRT systems resides in the way in which the self-organisation of information processing (coupled accumulation of template information and emergence of its interpretation) is dependent on the self-organisation of physico-chemical processes (emergence of macroscopic order as a result of reaction-diffusion coupling). The possibility of emergent spatio-temporal order in physico-chemical (thermodynamic) systems provided by Turing reaction-diffusion coupling bears within it the possibility of emergent molecular biological coding in GRT systems, including survival of the necessary reflexive information as a result of Darwinian selection. At first sight the connection between the two necessarily linked modes of self-organisation, pattern formation in reaction-diffusion systems and coding self-organisation in GRT systems, defies any simple mechanistic description. Nonetheless, the connection exists, and was carefully analysed by Füchslin and McCaskill (2001). The Turing mechanism operates essentially to colocalize templates with catalysts in a manner that establishes a codependency between them to the extent that their association defines them as belonging to a identifiably integrated system. The same effect has been achieved in other model systems, such as the autopoietic systems of Maturana and Varela (1980) or chemoton system of Gánti (2003), by endowing them with a membrane-forming capability for colocalizing molecular components. Autonomous creation of a system boundary allows the system as a whole to become the unit of selection, rather than the dynamic behaviour being dictated by internal competition for survival within or between classes of polymers. GRT systems are simpler in that the required spatial localization is given "for free" by the physically lawful system dynamics and does not have to be set up as an explicit feature of the model. Neither autopoietic or chemoton models demonstrate the emergence of coding or any other system for the interpretation of polymer sequences as meaningful information. They therefore lack a genotype-phenotype mapping of the sort characteristic of systems which undergo Darwinian selection as a result of genetic inheritance – in terms of von Neumann (1966) they do not carry a "description" of themselves and in terms of Füchslin and McCaskill (2001) they do not have a general system for the decoding (interpretation) of template information.[9] The process whereby systems of interpretation self-organise during biological evolution has been described elsewhere as *informed generation* (Wills, 2009).

Finally, it would be remiss not to point out that the Turing mechanism is only capable of mitigating the need for a minor miracle in relation to the emergence of the complex system of molecular biological coding observed in every living cell. Kick-starting the universal genetic code through Turing-type reaction-diffusion self-organisation would require simultaneous fluctuations that selectively increased the concentrations of the "correct" 20 out of 400 or more possible distinct assignment catalysts, along with a template bearing the corresponding reflexive information and correct coding for the replicase enzyme. Such an event is so improbable that its occurrence would qualify as another "miracle" which placed the main event of life's origin beyond any useful scientific analysis. However, the GRT systems in which self-organisation from an initially disordered state has been observed involve very modest amounts of template information, only the most rudimentary coding systems (one-bit codon and amino acid analogues) and simultaneous fluctuations in two out of four coding-assignment functions. When this is taken together with previous analysis of the hierarchy of nested constraints that would allow coding to emerge in a stepwise fashion, becoming increasingly complex and specific (Nieselt-Struwe and Wills, 1997), and the observation that the dynamics of coding self-organisation allow it to occur progressively (Wills, 2004; 2009), we see that the GRT model indeed provides a plausible explanation of the emergence of semiosis in prebiotic physico-chemical systems and the genotype–phenotype mapping necessary for biological.

---

[9] To give Gánti proper credit it must be added that he did not consider that a von Neumann description was necessary for biological inheritance and epigenetic effects prove him correct. However, it is hard to envisage how the exquisite control of processes in biological systems could be achieved without the logarithmic reduction in the complexity of the system's specification that an information-based mapping from cause to effect makes available.

The conclusions to be drawn from studies of GRT systems are:
  (i)   genetic information is created through the competitive process of natural selection;
  (ii)  functional interpreters construct themselves through a cooperative process of "informed generation": systemic self-organization of their component parts;
  (iii) the natural selection of apposite genetic information requires the operation of an interpreter;
  (iv)  the generative construction of a functional interpreter requires a store of apposite genetic information;
  (v)   the increasing complexity and specificity of biological forms requires the coevolution of genetic information and a system of functional interpretation.
The perspective on the emergence of genetic coding offered by GRT systems may also prove amenable to empirical investigation.

Empirical considerations

I now present a brief summary of an argument, to be developed in more detail in a separate paper (Wills & Nieselt, forthcoming), to the effect that the present day enzymes that serve the role of "translatases" in GRT systems are an example of a phenomenon that has a more plausible explanation based on semiotic principles than reductive physico-chemical reasoning supplemented with Darwin's principle.

Every cell contains a suite of 20 amino acyl-tRNA sythetase (AARS) enzymes, one for each of the 20 canonical amino acids from which proteins are made. The role of these enzymes is the same as the model translatases that catalyse codon to amino acid assignments in GRT systems. The origin of coding specificity is an apparent problem of the sort "you can't get here from there", requiring a "miracle" as a solution. An apparent paradox resides in the molecular biological inference that the specificity of codon-to-amino-acid assignments displayed by the AARSs depends on the specificity of their structures which cannot be maintained without the pre-existing specificity of their assignment functions (due to the essentially autocatalytic character of their production as an outcome of their interpreting reflexive sequence information in part of the cell's genome template). The life of every cell signals its existence in a thermodynamically privileged state that has arisen as a result of not only the naturally selected DNA template information it contains but also the highly organised constellation of molecular functionalities it inherited from its parent(s) (Wills, 2009). So, do cells hold any trace of this privileged state having bootstrapped itself into existence at life's root? It would seem so. The amino acid sequences of the AARS enzymes carry what can reasonably be construed as a palimpsest of (i) primordial digital codon and amino acid alphabets; and (ii) the structural core (Figure 1) of the corresponding molecular biological interpreter that executed an elementary binary code. This primæval binary code is hypothesized to have existed at a time which can be thought of as the origin of life, perhaps before anything remotely like an integrated single-cell organism had appeared on the planet. Whether an imprint of the entire pathway from an ancient binary code to the universal genetic code can be found in the data is still an open question, in fact the endpoint of coding evolution may not have been reached before different threads separated into the first semi-autonomous organisms. However, the unfolding of the genetic code in a tree of bifurcations in the amino acid recognition/differentiation process is clear in the inferred phylogenies of the AARSs (O'Donoghue and Luthey-Schulten, 2003; Fournier *et al.*, 2011) as well as transfer RNAs (Delarue, 2007).

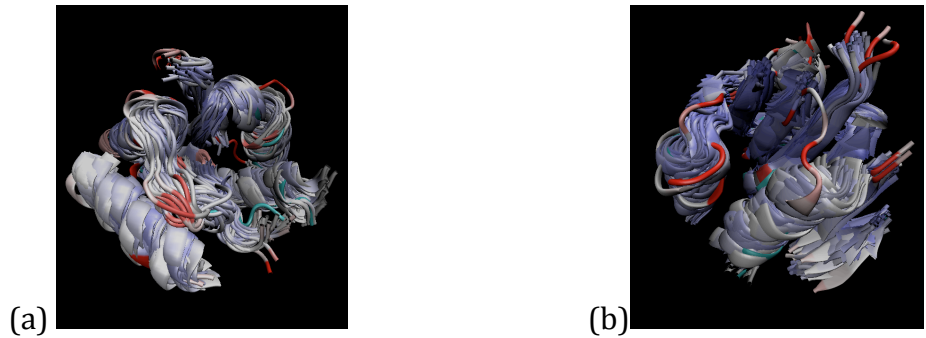

(a) (b)

Figure 1. Superimposed core structures (peptide α-carbon chains) of representative (a) Class I and (b) Class II AARS enzymes of all catalytic specificities from organisms belonging to diverse taxa from all biological kingdoms. Images were produced by using the VMD software (Humphrey *et al.*, 1996).

The suite of 20 AARS enzymes found in every cell is divided into two disjoint classes whose division is as universal as the genetic code. While the structure of all of the Class I enzymes from species belonging to all of life's kingdoms resemble one another, as do the Class II structures, there is nothing obvious in common between structures of enzymes from the two separate classes (Figure 1). On the other hand, both classes of AARSs fulfil the same generic biochemical function, catalysis of the attachment of an amino acid to a tRNA molecule. This chemical reaction is the process whereby the genetic code is executed. Code execution requires an AARS protein catalyst uniquely to recognise both an amino acid and a cognate tRNA and then attach them to one another. The accuracy with which this task is accomplished guarantees the fidelity of the universal genetic code.

Why do the cells of organisms from all branches of the tree of life, some of which diverged from one another nearly 4 billion years ago, all still carry versions of two solutions to a single biochemical problem, the ligation of an amino acid and tRNA molecule? The fundamental tenet of Darwinian evolution, survival of the fittest as a result of competition among variants generated in an error-prone production process, has been applied successfully to autocatalytic macromolecules (Eigen, 1971) and somewhat less successfully to single genes (Dawkins, 1976; Hubbard, 2013; Newman, 2013). This view of molecular evolution engenders the expectation that variation among and competition between the Class I and Class II solutions to the AARS problem would have led rapidly to the dominance of the gene for one and the demise of the gene for the other in the struggle for reproductive superiority. If the value of a primitive AARS to a system lay in its various capabilities of accurately discriminating and ligating specific amino acids and tRNAs, then differences in the fitness of prototypical Class I and Class II AARSs would have left a single survivor, not a pair of alternatives acting in concert. However, if entities as complex as organisms cannot exist without the prior invention of some means of meaningful computation, *i.e.*, the storage of information and the execution of a coded interpretation of it, then the cooperative survival of Class I and II AARS prototypes (Figure 1) can be viewed as a locked-in feature of molecular biological systems left over from the invention of the first-ever chemical code – cooperative survival of structurally dissimilar entities offered the only chemically feasible way of producing the two functional specificities $A \rightarrow a$ and $B \rightarrow b$ so that *neither* would be significantly confused with *either* $A \rightarrow b$ and $B \rightarrow a$ in the limited "bandwidth" available for the storage of genetic information in an error-prone system.

Thus, we infer that the first physico-chemical coding system in which information *per se* was created and simultaneously interpreted was as simple as possible, consisting of an elementary code linking two binary alphabets, one comprised of two distinguishable types of amino acids $\{a, b\}$ and the other comprised of two distinguishable types of nucleic acid codons $\{A, B\}$. The coded mapping from one binary alphabet to the other allowed information stored in nucleic acid sequences to be interpreted as proteins, whose collective catalytic functions corresponded to the restricted subset of codon-to-amino acid assignments $\{A \rightarrow a, B \rightarrow b\}$ comprising the code. There is no reason in principle why the enzymes catalysing these primitive assignments by means of tRNA amino-acylation should have been as vastly different from one another as the Class I and II AARSs,[10] but a system containing a structure that strongly preferred both *A* over *B* and *a* over *b* as substrates, and a completely different structure with complementary substrate preferences of equal strength, would more easily have staved off the corroding effects of the alternative coding assignments $\{A \rightarrow b, B \rightarrow a\}$ (Nieselt-Struwe and Wills, 1997). Furthermore, there is increasing evidence that the Class I and II core structures were encoded in the complementary strands of a single nucleic acid gene (Chandrasekaran *et al.*, 2013).

Concluding discussion

We began with the Shannon definition of *information*, which equates the information assigned to an object with a pattern that is distinguishable from other patterns in some arbitrarily defined class of possible properties of an equally arbitrarily defined class of objects. Shannon information can most simply be represented, and then measured, by using an alphabet of *symbols* (or letters), of which the simplest example is set of the binary digits $\{0, 1\}$. We then enquired into the *meaning* of any body of information and determined that absolutely any meaning is possible, any actual meaning depending only on the existence and operation of a physico-computational system called an *interpreter* that takes the information as input and produces the chosen meaning as output. This perspective was used to frame the fundamental problem of the origin of biological systems: what minimal conditions are needed for interpreters capable of assigning meaning to physical patterns, in particular, polymeric sequences, to emerge from disordered molecular systems? We found that the gene-replicase-translatase (GRT) model possesses all of the features needed satisfactorily to answer the question. In GRT systems genetic information is created through a process of *natural selection* in which the survival-value of information-carrying polymers is determined by their aptitude to serve as templates for the self-construction of a functional interpreter through a process of *informed generation*. The first molecular biological interpreter appears to have operated a simple *binary code*. Every living cell carries a palimpsest of the emergence of genetic coding, the surviving remnants of the primordial system of interpretation, in the dual structural cores of the Class I and II AARS enzymes. Life originated in the spontaneous creation of symbolic information and its interpretation. It was a result of thermodynamically driven self-organisation, the coupling of natural section and informed generation, in an autocatalytic system of nucleic acids and peptides.

---

[10] The bifurcation of the sets of codons and amino acids into functionally distinct subsets did not necessarily require the emergence of two structures as distinct as those portrayed in Figure 1, but that is what appears to have happened and become locked into all subsequent molecular biological evolution.

At the centre of every cell's operation there exists a pair of structures (Figure 1) whose forms have been preserved through every new generation in countless branches of life for nearly 4 billion years. These forms appear to have been the joint producers and products of the original elementary difference that was recognised and exploited by nature to create a bit of meaningful symbolic information. "A difference that makes a difference" (Bateson, 1972) emerged out of the sea of disordered chemical reactions and enslaved the flow of energy to serve the purpose of its own existence. The pragmatic distinctions made simultaneously between two classes of nucleic acid codons and two classes of amino acids allowed codon sequences to be taken as symbolic representations of the amino acid sequences of proteins which functioned to make the necessary distinctions. Semantic closure (Pattee, 1982; 1995) was achieved through a thermodynamically driven process of computational self-organisation and meaning was born locally on this planet. Now, you, the reader are looking at these words and an optical image of the letters is stimulating cells on the surface of your eyes' retinae and an astronomically more complex process of information processing is taking place as you register ("understand") the meaning of the words which I, the writer, have previously composed. The detailed process of communication and understanding is conditioned not only by the common structure of our participating brains, but the long process of cultural evolution that has resulted in the conventions that define the English language. All of this live information processing is sustained by and occurs in systems that incorporate in their most basic nanoscopic operating structures faithfully copied offspring of the molecules in which a bit of meaningful information first came into being.

Eigen (2013; p 449) claims to have answered the question of how complex interpretations of information originate by characterising the dynamics of biological evolution as movement in a high dimensional "I-space" of genetic information (Eigen 2013, pp354–387; 404–423), the very same space through which Dawkins (1986, pp72–73) believes an engineer could cobble a pathway from a pigeon to a dodo. Both of these authors seem to admit that information *per se* can function causally in physical systems: Dawkins (1986, p111) in his ecclesiastical proclamation of the "plain truth" that willow seeds are the equivalent to floppy discs; and Eigen (2013, p 480) when he says "In order for life to come about, there must be some physical principle that controls complexity" and goes on to ascribe the beginning of evolution to the origin of genetic information. According to Eigen (2013, p 479) we are in possession of the general law of which life is a consequence. Natural selection generates genetic information of which the apparently purposeful functional order of biological systems is a product. Neither Dawkins nor Eigen looks at the other side of the coin and engages seriously with the possibility that information is the product of the functional order of biological systems. Thus, Eigen (2013) devotes hundreds of pages to the task of explaining how natural selection generates information but, even though some relevant work has been conducted under his patronage (Hoffmann, 1974; 1975; Wills, 1993; Füchslin and McCaskill, 2001), makes no mention of processes whereby functionally ordered *interpretation* in biological systems makes the natural selection of genetic information *per se* possible. And when it comes to the genetic code, it is nucleic acids (tRNAs) rather than proteins (AARSs enzymes) that he claims "take care of the correct assignment of each of the 20 natural amino acids used in proteins to their cognate codons in messenger RNA (mRNA), as is necessary for error-free translation" (Eigen 2013, p 489). According to this view functional order in biological systems is not capable of generating anything fundamental, only epigenetic phenomena, which must be governed by the principle of natural selection and that principle alone.

In complete contrast with Dawkins (1986) and Eigen (2013) I have described how the epigenesis of molecular biological interpreters points to the kernel of nature's generative activity, an aspect of the physical universe as fundamental as those aspects embodied in the laws of physics and Darwin's principle. Of course the transition to a self-organised state in GRT-type systems or the more primitive quasi-species systems of Eigen (1971) could be described in a reductive physico-chemical narrative that didn't even acknowledge Darwin's natural selection as a principle. However, in spite of its validity, such a description would fail to signal the semi-autonomous role of information in biological systems (Schrödinger, 1944), without which genetic engineering would be impossible. By the same token, the narrative of Dawkins and Eigen fails to signal the semi-autonomous role of generative computation in the universe, the crucial distinguishing feature that separates biological systems from the other products of nature. It is necessary to combine the natural selection of polymer template sequences (Eigen, 1971) with the informed generation of interpreters (Wills, 2009) to explain how either information or function can come naturally to reside in molecules. In biological systems there is no information without function, or *vice versa*. The two co-evolve and neither has complete control.

The origin of semiosis in natural systems has been the subject of a great deal of philosophical argument, little of which has gained any significance from, or contributed any significance to, current theories of the elementary molecular properties of biological systems or features of protobiological systems. On the other hand, the main argument of this paper is that only by taking account of the semiotics of molecular biological coding, the origin of genetic information and its interpretation, can biological systems be investigated in a properly scientific manner. For more than half a century scientists have peddled the idea that DNA genes, selfish or otherwise, are the key to the understanding, and even more importantly now, the manipulation, of biological systems. Without doubt, DNA has served as a medium for information storage through æons of evolution. The base-paired structure of DNA, usually represented in its iconic double helical form, displays the aptitude of this polymer for this function – the complementary strands can be separated, allowing the sequence to be read and a new complementary copy to be written. However, DNA replication is what we have characterised as a "zeroth order" process of interpretation. Biology is replete with processes that confer much greater meaning on DNA sequence information. Correspondingly, the evolutionary maintenance and development of *meaning* and *interpretation* has been the domain of other special structures, created and sustained by various modes of thermodynamic self-organisation. The first non-trivial meaning conferred on molecular information seems to have been a capability of conferring meaning: the execution of a code to produce the entities with code-executive capabilities. Molecular biological self-construction and semiosis have a common origin in the emergence of the genetic code.

Semiosis requires not just physical stability of sign system instantiation (so that information can be inherited and serve as a codescript for the creation and maintenance of a daughter system), but also the dynamic stability of the system of interpretation (execution, utilization) of the information. The interpreter system of any biological species is a complex network of processes that has been built up through numerous symmetry breakings, phase transitions and self-organising "decision processes" (Wills, 2009). During reproduction its inheritance is as important as that of the genetic information. Inherited genetic information is useless in the absence of a functioning molecular biological interpreter. On the other hand genetic information serves as a fixed and necessary "boundary condition" or constraint (Pattee, 1972; 2013), without which the interpreter system would undergo thermodynamic collapse ("die").

Thus, we have characterised biological systems primarily as natural system which carry and interpret information. Their most important feature, their "life", is that they possess a codescript description of themselves (Schrödinger, 1944), which they are able to interpret (execute) as instructions for their own construction (von Neumann, 1966). Their physical structure embodies a system of correlation (genotype-to-phenotype mapping) whose existence requires the satisfaction of conditions (semantic closure) that can only be expressed in terms of extra-physical informational relationships. We have pointed to the origin of the genotype-phenotype mapping as the most important problem in theoretical biology. How does this orderly relationship, necessary for Darwinian selection as a result of variation in the inheritance of traits, arise? Our provisional answer: first in the emergence of a self-constructing system of genetic coding, and the self-representation of the code interpreter in genetic information; that is, in biosemiosis.

Endorsement

It is the author's wish that no agency should ever derive military or purely financial benefit from the publication of this paper. Authors who cite this work in support of their own are requested similarly to qualify the availability of their results.

Acknowledgments

I am grateful to Kay Nieselt for her hospitality in Tübingen and to the Carl Zeiss Foundation for financial support.